## CSS090530:144011+494734: a new SU UMa-type dwarf nova in Bootes

David Boyd, Nick Dunckel, Jerry Foote, Ian Miller

### Abstract

We report photometry and analysis of a previously unknown SU UMa-type dwarf nova in Bootes detected in outburst by the Catalina Real-time Transient Survey on 2009 May 30 with the discovery identifier CSS090530:144011+494734. This apparently stellar object had previously been catalogued by the Sloan Digital Sky survey as SDSS J144011.01+494733.4. We measured its mean superhump period over the first 3 days following detection as 0.06500(4) d at which point this changed to 0.06438(3) d. We detected a possible orbital period of 0.06322(8) d implying a mean superhump period excess of 0.020(2). After detection, the dwarf nova maintained steady brightness for 5 days before starting to fade. The outburst amplitude was 5.5 magnitudes above its quiescent level.

### Discovery

A new transient object in Bootes was observed at magnitude 15.7 by the Catalina Real-time Transient Survey (CRTS) [1] on an image taken on 2009 May 30 at 07.20 UT [2] and was confirmed on three further images over the next 27 minutes. CRTS assigned it the discovery identifier CSS090530:144011+494734. It was listed on the CRTS webpage of detected cataclysmic variables [3] and included in a list of variable star candidates published in ATEL #2086 [4]. The discovery was subsequently reported on VSNET by Kato on May 31 at 02.00 UT [5]. The position was reported in ATEL #2086 as RA 14h 40m 11.04s Dec +49° 47' 33.8" (J2000). There is a faint object at that position on the Digitized Sky Survey POSS1 and POSS2 blue plates [6] and a blue, apparently stellar, object was recorded at that position by the Sloan Digital Sky Survey (SDSS) [7] with identifier SDSS J144011.01+494733.4 and magnitudes g = 21.18 and r = 21.01. This object has not previously been identified as a cataclysmic variable candidate in SDSS publications [8–13]. Previous CRTS images taken on 13 occasions since 2006 with a limiting V magnitude of ~19.5 did not record an object at that position. Neither has anything been detected by the All Sky Automated Survey [14] above a limiting V magnitude of about 14 or by the Northern Sky Variability Survey [15] above a limiting magnitude of about 15.5. There is no object listed in Simbad [16] within 10 arcsec of this position and the only coincident entry in Vizier [17] at the time of the outburst was for the SDSS object noted above. An entry for the new variable was subsequently created in the AAVSO International Variable Star Index (VSX) entry [18].

### Observations

Observations by the authors commenced on May 31 at 22.30 UT and unfiltered time-series totalling 41 hours plus individual magnitude measurements were obtained on subsequent nights until the object was too faint to measure reliably. Superhumps were immediately evident confirming that this was a SU UMa-type dwarf nova observed in outburst for the first time. A log of observations is given in Table 1 and the equipment used is listed in Table 2.

V and R magnitudes for possible comparison stars were calculated from r' magnitudes in the CMC-14 catalogue and J and K magnitudes in the 2MASS catalogue (both accessible through Vizier) using the formulae of Dymock and Miles [19] and Greaves [20]. Comparison stars were selected which had relatively blue colour and which showed no significant variation. These are listed in Table 3 along with the magnitudes adopted. An image of the field taken on June 1 is shown in Figure 1 indicating the new variable at magnitude 15.8 and the comparison stars listed in Table 3. Also in Figure 1 is an image of the same field taken on June 12 when the variable had faded to magnitude 19.1. The variable is the SW component of a close pair.

All images were dark-subtracted and flat-fielded and instrumental magnitudes in each image measured using aperture photometry. Absolute magnitudes of the variable were then calculated with reference to one or more of the comparison stars. Magnitude measurements by different observers which overlapped in time were found to differ by up to 0.15 magnitude because of different choices of comparison stars and different spectral responses of their equipment. The relative magnitude differences between observers were found by comparing their light curves in regions where these overlapped and the magnitudes of all observers were brought into mutual consistency by increasing or decreasing their magnitudes by a constant amount. All measurement times were converted to Heliocentric Julian Dates. The resulting light curve of the outburst is shown in Figure 2 with the discovery measurement marked as a cross. The dwarf nova remained at a similar brightness for 5 days following detection before fading gradually back towards quiescence. If we take the SDSS g magnitude of 21.2 as its quiescent level, the outburst amplitude was 5.5 magnitudes.

**Astrometry**

Using Astrometrica [21] and the UCAC2 catalogue, the mean position of the new variable was measured from 8 images taken on 2009 June 1 as RA 14h 40m 11.00+/-0.01s Dec +49° 47' 33.5+/-0.1" (J2000). This is in reasonable agreement with the position given by CRTS and in good agreement with the SDSS position. The separation between the variable and the nearby star to the NE is 6 arcsec.

**Superhump period analysis**

Twenty-one times of maximum for 17 separate superhumps were found by fitting a 2$^{nd}$ order polynomial to the light curves in the region of each maximum. A further 3 times of maximum from Japanese observers were obtained from [22]. Superhump cycles numbers were assigned by inspection and a weighted linear fit made to all times of superhump maximum. This gave a mean superhump period during the outburst of 0.064459(9) d. Observed minus calculated (O-C) times of maximum relative to this linear fit are plotted in Figure 3. Our data are shown as filled circles, those from [22] as crosses. In this plot, regions of constant superhump period will be represented by straight lines whose gradient gives the period. There appear to be two discrete regions before and after cycle 30 (HJD = 2454985.0) with approximately constant superhump periods in each region and with a period change around cycle 30. Weighted linear fits to the times of superhump maximum before and after cycle 30 are shown as dotted lines in Figure 3. The mean superhump period before cycle 30 is $P_{sh}$ = 0.06503(9) d and after cycle 30 is $P_{sh}$ = 0.06437(2) d. O-C times of maximum relative to these linear fits are listed in Table 4 and 5 respectively.

As the data are relatively sparse, more complex interpretations are also possible. In particular, reasonable quadratic fits can be applied separately to the data before and after cycle 50. The fit after cycle 50 implies an increasing period during this region with a rate of period change dP/dt = 14(6) * $10^{-5}$. Although this is a relatively large positive rate of change, it is nevertheless possible (see Figure 10 in [22]). We also investigated a quadratic fit to the times of maximum for all the data which would have implied a steady decrease in the period throughout this interval but this gave a poor fit.

After subtracting mean magnitudes from each run, separate period analyses of the data before and after cycle 30 were performed using the Lomb-Scargle method [23, 24] in Peranso [25]. These gave the power spectra shown in Figure 4. The strongest signals in each case are due to the presence of superhumps. The superhump period obtained from period analysis before cycle 30 was $P_{sh}$ = 0.0649(3) d and after cycle 30 was $P_{sh}$ = 0.0645(1) d. Both power spectra show numerous alias signals, those before cycle 30 being at 1 c/d difference in frequency as expected since these observations were all made from essentially the same longitude. After cycle 30 the alias structure is more complicated as these measurements were from widely separated locations. Superhump phase diagrams obtained by folding the data before and after cycle 30 on these two periods are shown in Figure 5. The amplitudes of the superhump signals before and after cycles 30 are 0.16 and 0.10 magnitude. Weighted mean superhump periods before and after cycle 30 calculated from the results of superhump timing analysis and period analysis are given in Table 6. A period analysis of all the data taken together gave a mean superhump period of $P_{sh}$ = 0.06447(8) d.

In each of these two regions the superhump signal was removed and the period analysis repeated giving the power spectra shown in Figure 6. The periods of the strongest residual signals before and after cycle 30 were 0.0633(3) d and 0.0632(2) d respectively, although the latter signal is rather weak. Removing these signals left only low level signals none of which had periods with a significant relationship to any of the signals which had been removed. The similarity of these two residual signals suggests that they may represent the orbital period. Although we have only circumstantial evidence for this, we tentatively adopt the weighted mean of these two signals $P_{orb}$ = 0.06322(8) d (1 hr 31 min) as the orbital period of this binary system. The orbital phase diagram obtained by folding all the data on this period is shown in Figure 7. The mean amplitude of this signal is 0.035 magnitude.

Using the mean superhump and assumed orbital periods, the mean superhump period excess $\varepsilon$ is 0.020(2), which is consistent with results for other UGSU dwarf novae with similar orbital periods [22, 26]. If we assume this is a normal dwarf nova, then using the empirical relationship $\varepsilon = 0.18*q + 0.29*q^2$ between superhump period excess and mass ratio q given in [27], we obtain a secondary to primary mass ratio for the binary system of 0.095(8).

## Conclusions

Photometry of a new transient object in Bootes discovered by the Catalina Real-time Transient Survey on 2009 May 30 has revealed it to be a previously unknown SU UMa-type dwarf nova. We measured superhumps with a mean period of 0.06500(4) d during the first 3 days of observation at which point their period changed to 0.06438(3) d. We detected a possible orbital signal at 0.06322(8) d implying a mean superhump period excess of 0.020(2), similar to other UGSU dwarf novae with this orbital period. Using a published empirical relationship, this gives a mass ratio for the binary system of 0.095(8). The dwarf nova maintained steady brightness for 5 days after initial detection before starting to fade. The outburst amplitude above quiescence was 5.5 magnitudes.

## Acknowledgements

We acknowledge with thanks the Catalina Real-time Transient Survey for making their real-time detection of transients available to the public and the Simbad and VizieR services operated by CDS Strasbourg. We are also grateful for helpful comments by the referee.

## Addresses

DB: 5 Silver Lane, West Challow, Wantage, Oxon, OX12 9TX, UK [drsboyd@dsl.pipex.com]
ND: 12971 Cortez Lane, Los Altos Hills, CA 94022, USA [ndunckel@earthlink.net]
JF: CBA Utah, 4175 East Red Cliffs Drive, Kanah, UT 84741, USA [jfoote@scopecraft.com]
IM: Furzehill House, Ilston, Swansea, SA2 7LE, UK [furzehillobservatory@hotmail.com]

| Start time (JD) | Duration (hrs) | Observer |
|---|---|---|
| 2454983.43768 | 3.40 | Boyd |
| 2454984.42093 | 2.96 | Boyd |
| 2454984.43476 | 3.83 | Miller |
| 2454985.42344 | 3.75 | Miller |
| 2454986.43483 | 2.60 | Boyd |
| 2454987.44188 | 3.44 | Miller |
| 2454990.43883 | 0.24 | Boyd |
| 2454990.67379 | 5.33 | Foote |
| 2454990.74683 | 4.31 | Dunckel |
| 2454991.69532 | 3.95 | Dunckel |
| 2454992.42780 | 0.20 | Miller |
| 2454993.44212 | 0.34 | Miller |
| 2454993.74848 | 0.44 | Dunckel |
| 2454994.42916 | 0.24 | Boyd |
| 2454994.70520 | 5.84 | Dunckel |

Table 1. Log of observations.

| Observer | Equipment used |
|---|---|
| Boyd | 0.35-m f/5.3 SCT + SXV-H9 CCD |
| Dunckel | 0.32-m f/9 Ritchey-Chretien + SBIG ST-10E CCD |
| Foote | 0.60-m f/3.4 reflector + ST-8e CCD |
| Miller | 0.35-m f/10 SCT + SXV-H16 CCD |

Table 2. Equipment used.

| Label (Figure 1) | Identifier | V | (V-R) |
|---|---|---|---|
| C1 | GSC 3476 454 | 12.91 | 0.411 |
| C2 | GSC 3476 640 | 14.30 | 0.526 |
| C3 | GSC 3476 780 | 15.43 | 0.302 |
| C4 | GSC 3476 462 | 15.47 | 0.467 |

Table 3. Comparison stars used.

| Superhump cycle no | Observed time of maximum (HJD) | Uncertainty (day) | O-C (day) |
|---|---|---|---|
| 0 | 2454983.02380 | 0.00310 | -0.00235 |
| 7 | 2454983.47650 | 0.00313 | -0.00484 |
| 8 | 2454983.54108 | 0.00284 | -0.00529 |
| 15 | 2454984.00430 | 0.00160 | 0.00274 |
| 16 | 2454984.06890 | 0.00110 | 0.00232 |
| 22 | 2454984.45655 | 0.00063 | -0.00019 |
| 22 | 2454984.45768 | 0.00122 | 0.00093 |
| 23 | 2454984.52120 | 0.00119 | -0.00057 |
| 23 | 2454984.52165 | 0.00201 | -0.00012 |
| 24 | 2454984.58573 | 0.00091 | -0.00107 |

Table 4. Times of superhump maximum and O-C values relative to a linear fit before cycle 30.

| Superhump cycle no | Observed time of maximum (HJD) | Uncertainty (day) | O-C (day) |
|---|---|---|---|
| 38 | 2454985.49293 | 0.00101 | -0.00040 |
| 39 | 2454985.55620 | 0.00157 | -0.00150 |
| 53 | 2454986.45988 | 0.00112 | 0.00101 |
| 54 | 2454986.52533 | 0.00124 | 0.00209 |
| 69 | 2454987.48473 | 0.00193 | -0.00404 |
| 70 | 2454987.55267 | 0.00273 | -0.00047 |
| 119 | 2454990.70611 | 0.00099 | -0.00111 |
| 120 | 2454990.77398 | 0.00520 | 0.00240 |
| 120 | 2454990.77562 | 0.00274 | 0.00403 |
| 121 | 2454990.83577 | 0.00297 | -0.00019 |
| 121 | 2454990.83848 | 0.00266 | 0.00252 |
| 122 | 2454990.90085 | 0.00334 | 0.00052 |
| 135 | 2454991.73678 | 0.01132 | -0.00034 |
| 136 | 2454991.80481 | 0.00540 | 0.00332 |

Table 5. Times of superhump maximum and O-C values relative to a linear fit after cycle 30.

| Superhump cycle range | $P_{sh}$ from superhump timing analysis | $P_{sh}$ from Lomb-Scargle period analysis | Weighted mean $P_{sh}$ |
|---|---|---|---|
| 0 - 30 | 0.06503(9) d | 0.0649(3) d | 0.06500(4) d |
| 31-150 | 0.06437(2) d | 0.0645(1) d | 0.06438(3) d |

Table 6. Mean values of $P_{sh}$ before and after cycle 30.

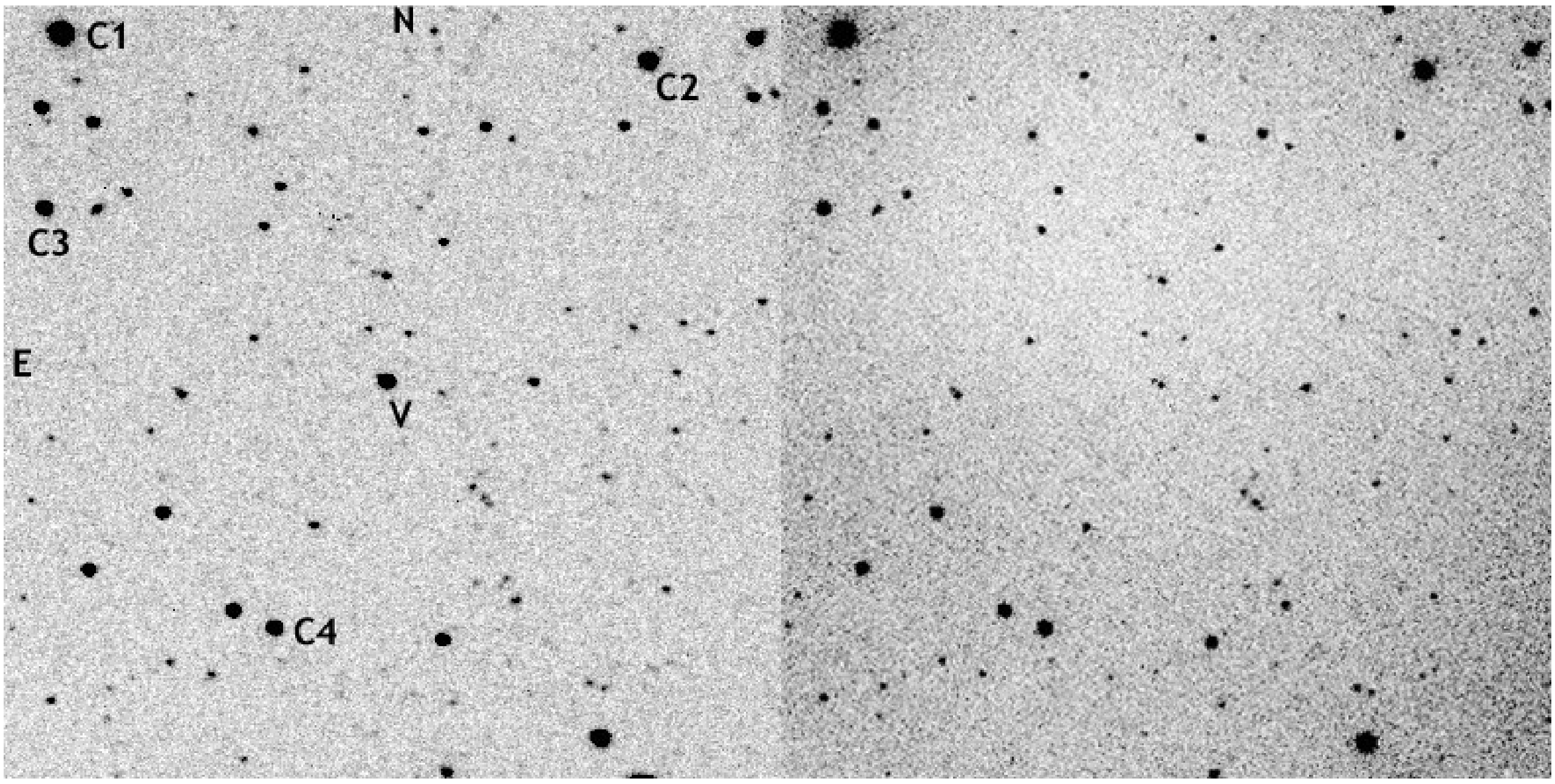

Figure 1. Images of the new variable (left) on June 1 at mag 15.8 with comparison stars marked (*Boyd*) and (right) on June 12 at mag 19.1 (*Dunckel*). Field approx 10' by 10'.

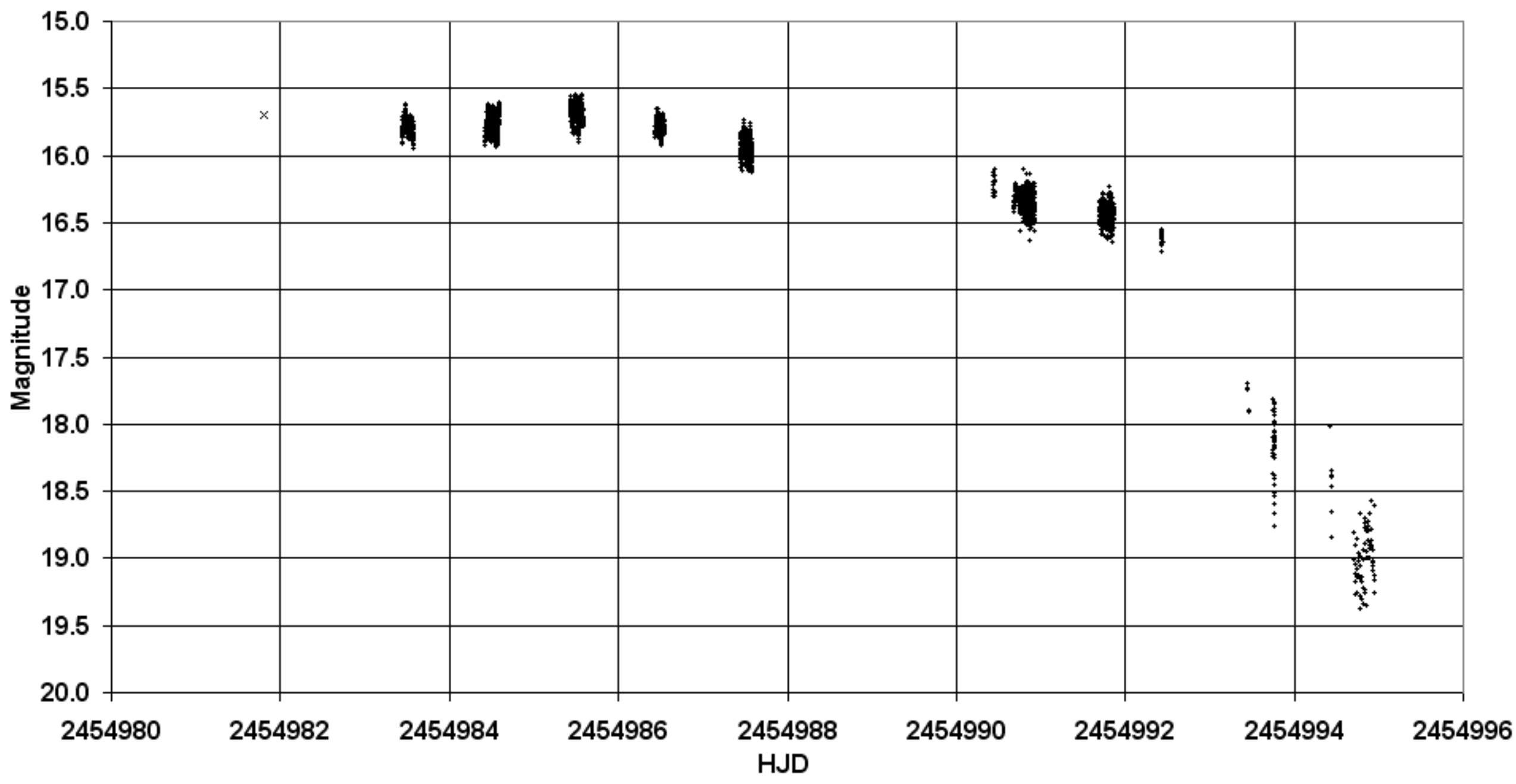

Figure 2. Light curve of the outburst.

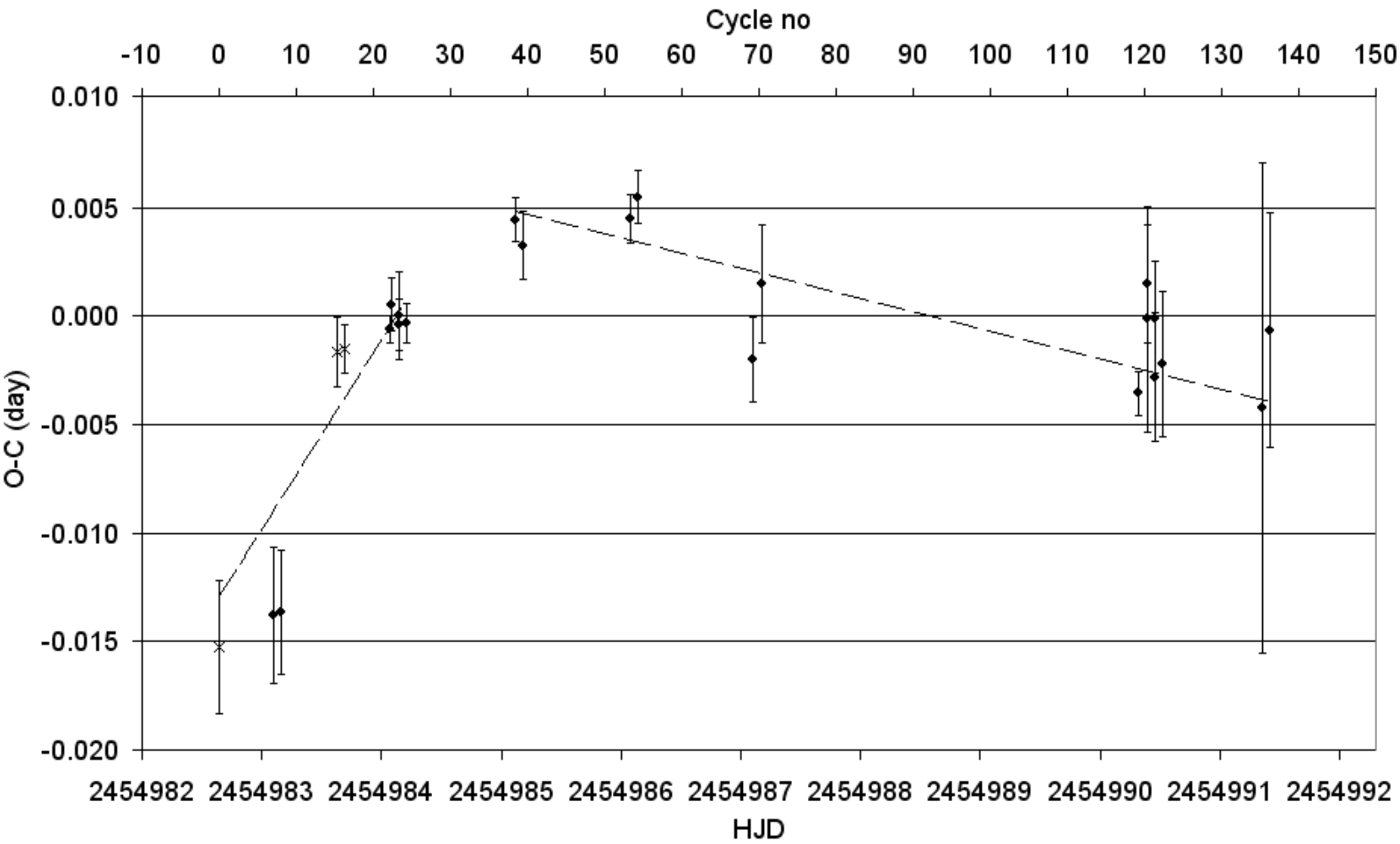

Figure 3. O-C times of superhump maximum relative to a linear fit to all times of maximum with weighted linear fits before and after cycle 30 shown dotted.

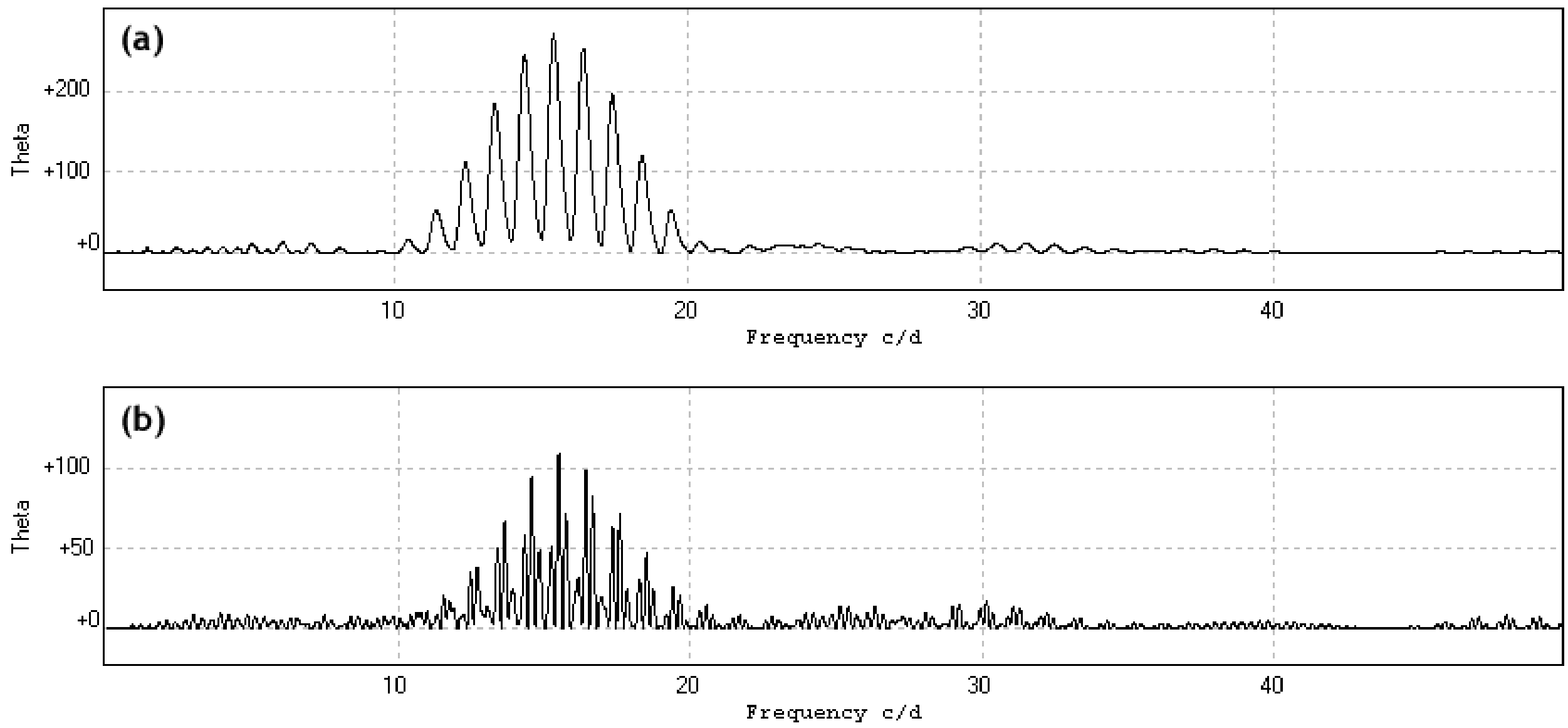

Figure 4. Power spectra from Lomb-Scargle analysis of data (a) before cycle 30 and (b) after cycle 30.

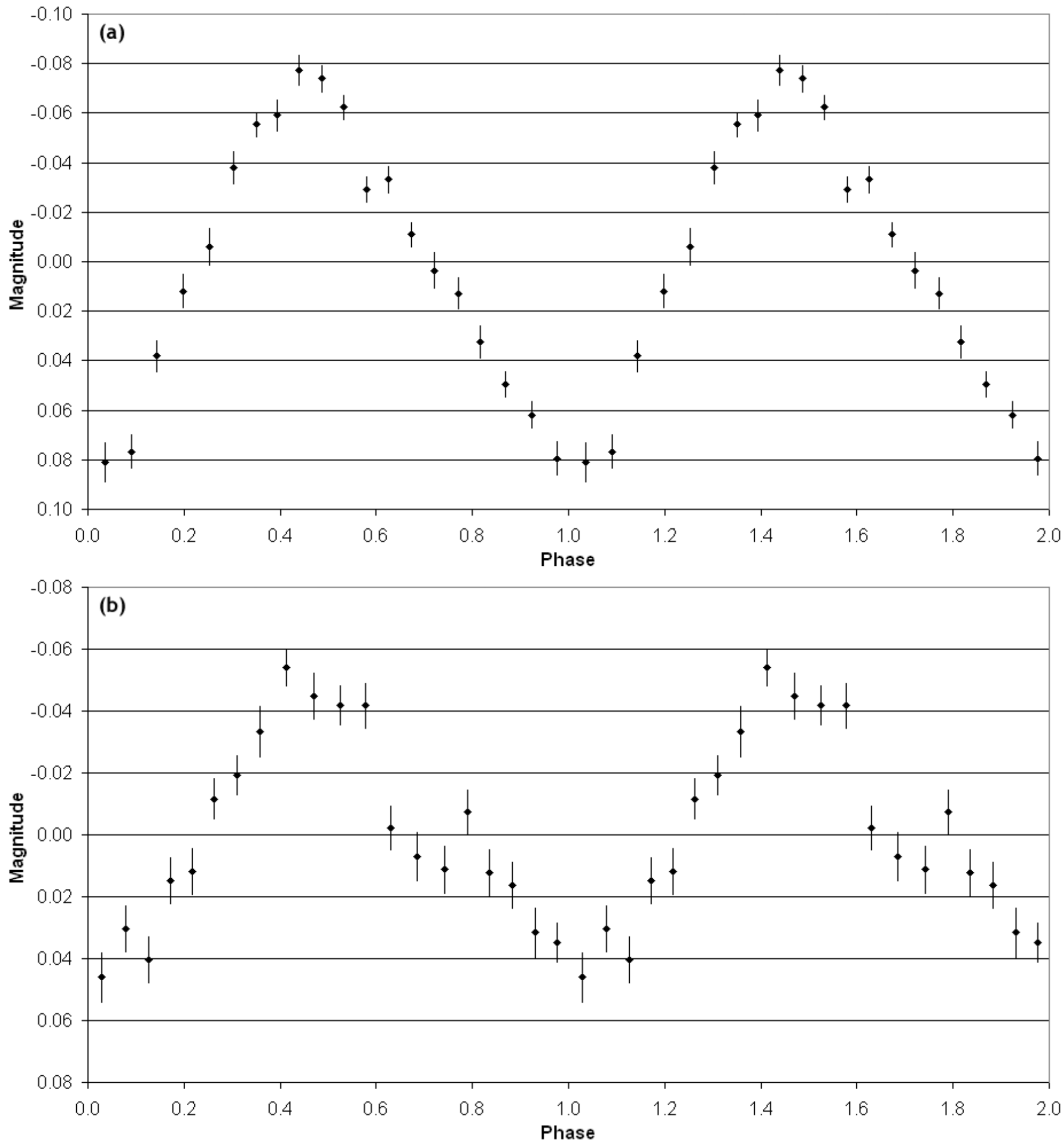

Figure 5. Superhump phase diagrams for data (a) before cycle 30 folded on 0.0649 d and (b) after cycle 30 folded on 0.0645 d (two cycles).

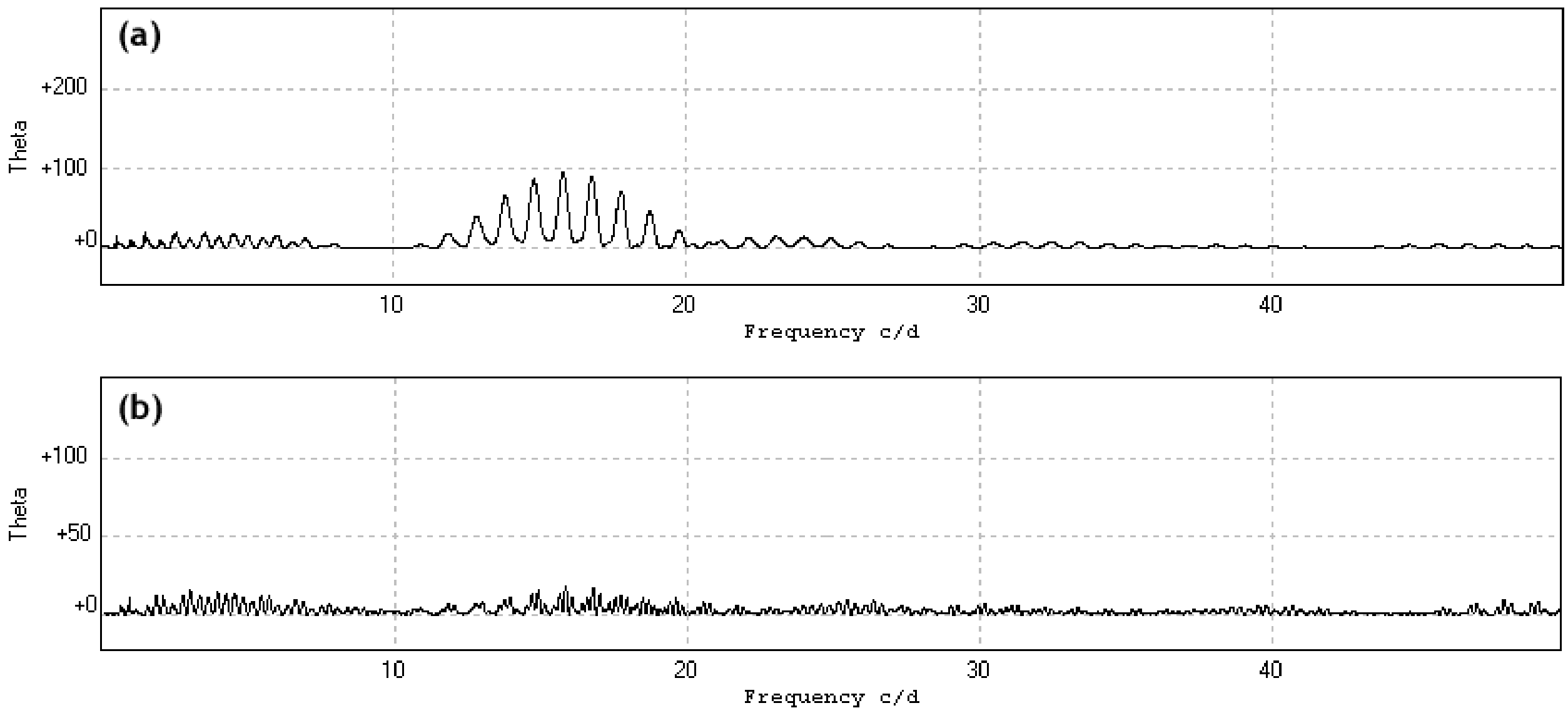

Figure 6. Power spectra from Lomb-Scargle analysis of data after removing the superhump signals (a) before cycle 30 and (b) after cycle 30.

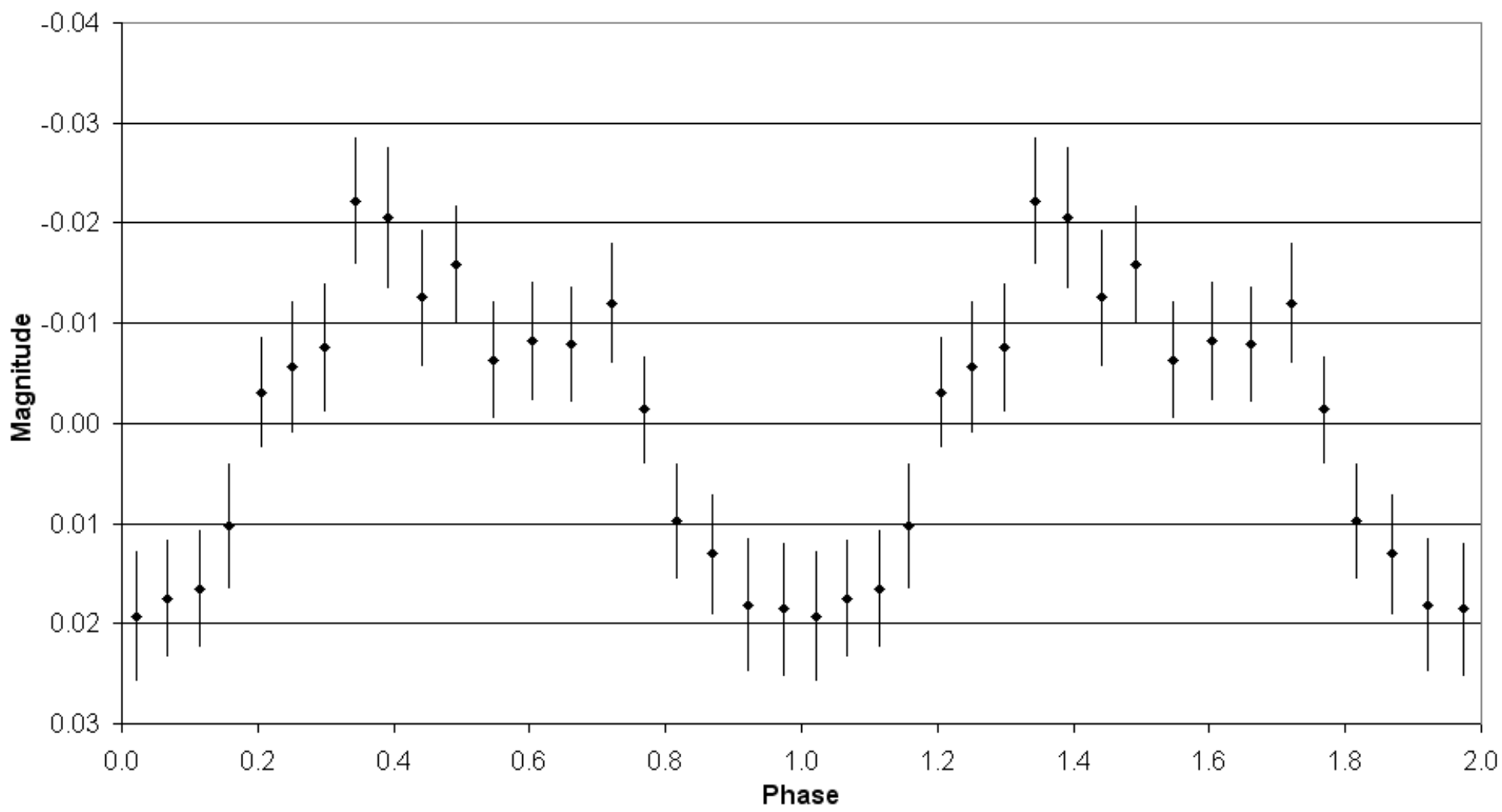

Figure 7. Orbital phase diagram for all data folded on 0.06322 d (two cycles).